\documentclass{article}
\usepackage{spconf,amsmath,graphicx,hyperref}

\usepackage{booktabs}  
\usepackage{multirow}
\usepackage{array}     
\usepackage[numbers]{natbib}
\usepackage{xcolor}
\usepackage{booktabs}
\usepackage{pgfplots}
\pgfplotsset{compat=1.18}
\usepackage{tikz}
\usepackage{graphicx}
\usetikzlibrary{positioning,calc,decorations.pathreplacing,arrows.meta}
\usepackage{todonotes}
\usepackage{spconf,amsmath,graphicx,hyperref}
\usepackage{amssymb}
\usepackage{amsmath}
\usepackage[table]{xcolor}
\definecolor{stampcolor}{RGB}{235,243,250}

\DeclareRobustCommand{\stamp}{\textsc{STAM-ASR}}

\title{\stamp{}: Speaker-Temporal Anchoring with Memory for Multi-Speaker ASR}

\name{Victor Tolulope Olufemi$^{*}$, Syeda Faiza Ahmed Sara$^{*}$, Shammur Absar Chowdhury
\thanks{$^{*}$Equal Contribution.}}

\address{Qatar Computing Research Institute (QCRI), Doha, Qatar\\}

\begin{document}
\ninept
\maketitle
%
\begin{abstract}
Natural conversations make both speech recognition and speaker attribution challenging for ASR, as speakers take turns, overlap, and reappear over time.
We propose \textbf{\stamp{}}, Speaker-Temporal Anchoring with Memory, a lightweight framework that extends an already pretrained AudioLLM for multi-speaker ASR. Without relying on an external diarization system, \stamp{} learns speaker activity and speaker-aware representations directly from intermediate AudioLLM features. Hence providing explicit \emph{who} and \emph{when} cues to modulate the AudioLLM's semantic representation without explicit speech separation. \stamp{} further maintains fixed-size speaker and conversational memories to carry complementary context across turns. We evaluate \stamp{} on AMI, ICSI, LibriCSS, and NOTSOFAR-1 across close-talk, far-field, overlapping, and cross-domain conditions. Our reported results shows that speaker-temporal conditioning and memory provide complementary benefits, while the gap between reference and
predicted speaker activity identifies robust speaker tracking as a key remaining challenge.
\end{abstract}

\begin{keywords}
Multi-speaker ASR, AudioLLM, Speaker-temporal modeling, Memory-augmented ASR
\end{keywords}
\section{Introduction}
\label{sec:intro}

Human conversation is dynamic where multiple speakers take turns, overlaps to acknowledge or interrupt one another among others.
To handle such diverse phenomena, multi-speaker ASR needs to recognize not only \emph{what} was said, but also \emph{who} spoke and \emph{when}.

Over the years, multi-speaker ASR has been studied from several directions, including speaker attribution, overlapping speech, and long-context modeling. Speaker-attributed ASR extends speech recognition beyond \emph{what} was said to also determine \emph{who} spoke, typically by combining recognition with diarization or target-speaker conditioning~\cite{kanda20_interspeech,kanda2022transcribe,delcroix2018single,polok2026dicow}. Whereas, multi-talker systems further address concurrent speech through continuous speech separation~\cite{chen2020continuous}, permutation-invariant recognition~\cite{taherian2022multi}, or serialized output training~\cite{kanda2020serialized,kanda2022streaming,raj2023surt}.

More recently, AudioLLMs have incorporated speaker selection, registration, and temporal grounding into language-model-based ASR~\cite{shi2024advancing,meng2025large,yin2026speakerlm,huo2026tagspeech,lin2026speakerreasoner}. For long recordings, speaker caches can maintain speaker identity across
chunks~\cite{shi2026train,peng2026g}, while long-context models propagate preceding utterances or compact recurrent states~\cite{hori20_interspeech,dai2019transformer,bulatov2022recurrent}. These approaches, however, largely address speaker tracking and contextual history separately, leaving open how to preserve both speaker-specific and conversational information across turns. 

To address this, we propose \textbf{\stamp{}}, \emph{Speaker-Temporal Anchoring with Memory}, a lightweight extension of a pretrained Omni AudioLLM for multi-speaker ASR. \stamp{} combines current speaker-temporal evidence with two fixed-size recurrent memories -- a \emph{speaker memory} that carries
acoustic-semantic information across appearances of the same speaker, and a \emph{conversational memory} that carries conversational context. Together, they preserve complementary speaker-specific and conversational information. 

Rather than introducing a separate acoustic frontend, \stamp{} reuses representations from the pretrained audio encoder. A lightweight internal diarization module estimates speaker activity and speaker representations, which provide explicit \emph{who} and \emph{when} cues to modulate the AudioLLM's semantic representation. This produces a speaker-conditioned view of the mixed audio without external diarization or explicit waveform separation.



\begin{figure}[t]
    \centering
    \scalebox{0.5}{
    \includegraphics[width=0.8\textwidth]{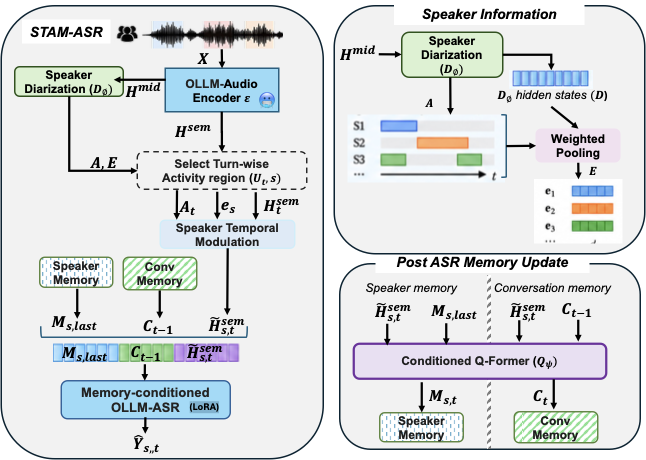}}
    \vspace{-0.2cm}
    \caption{Overview of \stamp{} with internal diarization, speaker-temporal modulation, memory construction. }
    \vspace{-0.5cm}
    \label{fig:overview}
\end{figure}

Therefore, our main contributions are as follows:
\textit{(a)} we introduce \textbf{\stamp{}}, which integrates speaker-temporal conditioning and memory into a pretrained AudioLLM for multi-speaker ASR without explicit speech separation; and
\textit{(b)} introduce separate fixed-size speaker and conversational memories to preserve complementary speaker-specific and conversation-level context across turns.

We evaluate \stamp{} across close-talk, far-field, overlapping, and unseen cross-domain conditions. Our analysis shows that its components play complementary roles: speaker-temporal conditioning is more effective when current speaker evidence is important, whereas memory provides additional benefit in meeting conditions where speaker and conversational history is informative.

\section{\stamp{}}
\label{sec:method}




We propose \textbf{\stamp{}}, Speaker-Temporal Anchoring with Memory, a lightweight extension of an AudioLLM for multi-speaker ASR. The key idea is to preserve the AudioLLM's acoustic-semantic representation while making it
speaker-aware. For each target speaker, \stamp{} uses speaker representation and temporal activity to modulate the AudioLLM representation without changing its sequence length or dimensionality. A Q-Former maintains fixed-size speaker and conversational memories that provide context across turns.
The framework consists of four components: (i) audio representation extraction, (ii) speaker diarization and representation pooling, (iii) speaker-temporal modulation, and (iv) memory-conditioned ASR.

\subsection{Audio Representation}
\label{sec:audio-representation}

Given audio $X$, the frozen AudioLLM encoder $\mathcal{E}$ extracts an intermediate representation $H^{\mathrm{mid}}$ for speaker modeling and a semantic representation $H^{\mathrm{sem}}$ for ASR. For long recordings, these representations are extracted over shorter audio segments, while speaker and memory states are propagated across segments.


\subsection{Speaker Diarization and Temporal Anchoring}
\label{sec:diarization}
We build the diarization module directly on $H^{\mathrm{mid}}$, avoiding a separate acoustic frontend. A Transformer-based diarization module $\mathcal{D}_{\phi}$ produces multi-label speaker activity $A$ and speaker-aware hidden states $D$, ($(A,D)=\mathcal{D}_{\phi}\!\left(H^{\mathrm{mid}}\right)$).
As $A$ is multi-label, multiple speakers may be active simultaneously, allowing overlapping speech to be represented explicitly.

We derive a representation for each speaker from the diarization hidden states using activity-weighted pooling,
\begin{equation}
\mathbf{e}_s =
\frac{
\sum_{\tau} w_{s,\tau}D_{\tau}
}{
\sum_{\tau} w_{s,\tau}+\epsilon
},
\end{equation}
where $\tau$ indexes diarization frames and $w_{s,\tau}$ is derived from the activity of speaker $s$. By default, $w_{s,\tau}=A_{s,\tau}$. 

The diarization module produces speaker activity and speaker-aware representations within each processed segment. For long recordings, speaker representations are associated across consecutive segments to
maintain consistent speaker identities.
We represent the $t$-th turn as $U_t=(s,a_t,b_t)$,
where $s$ is the diarized speaker identity and $a_t$ and $b_t$ denote the start and end timestamps. These timestamps are mapped to the corresponding indices of the precomputed AudioLLM and diarization representations to obtain the turn-level acoustic-semantic representation $H_t^{\mathrm{sem}}$ and target-speaker activity $A_{s,t}$.

\subsection{Speaker-Temporal Modulation}
\label{sec:speaker-modulation}

For each turn $U_t$ and target speaker $s$, \stamp{} conditions the
AudioLLM representation using two complementary cues: the speaker
representation $\mathbf{e}_s$ indicates \emph{who} is speaking, while
the activity $A_{s,t}$ indicates \emph{when} that speaker is active.
A lightweight conditioning network predicts feature-wise scale and shift
parameters, $(\gamma_{s,t},\beta_{s,t}) = f\!\left(\mathbf{e}_s,A_{s,t}\right)$.

Following we inject the speaker-conditioned information through a residual FiLM transformation~\cite{perez2018film},
\begin{equation}
\widetilde{H}_{s,t}^{\mathrm{sem}}
=
H_t^{\mathrm{sem}}
+
\lambda
\left(
\gamma_{s,t}\odot H_t^{\mathrm{sem}}
+
\beta_{s,t}
\right),
\end{equation}
where $\lambda$ is a learned scaling parameter. This preserves the sequence
length and dimensionality of $H_t^{\mathrm{sem}}$ while producing a
speaker-specific representation for ASR and memory updates.

\paragraph*{Overlap handling.}
When speakers overlap, they share the same mixed
$H_t^{\mathrm{sem}}$. \stamp{} conditions this representation separately
for each active speaker using its own $(\mathbf{e}_s,A_{s,t})$, producing
a distinct $\widetilde{H}_{s,t}^{\mathrm{sem}}$ without explicit waveform
separation.

\subsection{Memory-Conditioned ASR}
\label{sec:memory-asr}

\stamp{} maintains two fixed-size recurrent memories using a shared Q-Former $\mathcal{Q}_{\psi}$. Speaker memory follows successive turns of the same speaker, $M_{s,t}=\mathcal{Q}_{\psi}
(M_{s,\mathrm{last}},\widetilde H_{s,t}^{\mathrm{sem}})$,
while conversational memory follows chronologically turns, $C_t=\mathcal{Q}_{\psi}(C_{t-1},\widetilde H_{s,t}^{\mathrm{sem}})$.
Both contain a fixed number $N_{\mathrm{mem}}$ of tokens; a shared
Q-Former with memory-type embeddings distinguishes the two updates.


For ASR, the current turn uses the memory states available \emph{before} their update. Speaker and conversational memories are enclosed by fixed boundary tokens,
\begin{equation}
Z_{s,t} =
[
\langle\mathrm{sm}\rangle
M_{s,\mathrm{last}}
\langle/\mathrm{sm}\rangle;
\langle\mathrm{cm}\rangle
C_{t-1}
\langle/\mathrm{cm}\rangle;
\widetilde{H}_{s,t}^{\mathrm{sem}}
].
\end{equation}
Following, the AudioLLM then generates the corresponding speaker-attributed transcript, $\hat{Y}_{s,t} =\operatorname{Omni}\!\left(Z_{s,t}\right)$.

After transcription, both memories are updated for subsequent turns. We use a learned initial memory tokens, when no previous turns are observed.

\section{Experimental Setup}
\label{sec:experiments}

\begin{table}[t]
\centering
\caption{Corpora used for training and evaluation. 
Train/Dev/Test and A and D represent ASR and diarization supervision (Sup.).}
\label{tab:corpora}
\scriptsize
\renewcommand{\arraystretch}{0.88}

\begin{tabular*}{\columnwidth}{
@{\extracolsep{\fill}}lcccc@{}
}
\toprule
\textbf{Corpus} & \textbf{Sessions} & \textbf{Hours} &
\textbf{Spk.} & \textbf{Train Sup.} \\
\midrule
AMI-IHM
& 136/18/16 & 79.1/9.4/8.9 & 3--5/4/3--4 & A,D \\

ICSI
& 49/6/5 & 45.6/5.0/5.4 & 3--8/5--8/5--10 & A,D \\

Mixer6
& 189/--/-- & 51.2/--/-- & 1/--/-- & A \\

LibriSpeech-Sim.
& 5.2k/--/-- & 130/--/-- & 3--5/--/-- & D \\

AMI-SDM
& --/--/16 & --/--/8.9 & --/--/3--4 & A,D \\

LibriCSS
& --/--/60 & --/--/10.1 & --/--/8 & A,D \\

NOTSOFAR-1
& --/36/129 & --/3.7/13.2 & --/5--7/3--7 & A,D \\
\bottomrule
\end{tabular*}

\vspace{-0.2cm}
\end{table}

\subsection{Data and Evaluation}
\label{sec:data}

\textbf{Training:} Table~\ref{tab:corpora} summarizes the corpora. We train \stamp{} on
AMI~\cite{carletta2005ami}, ICSI~\cite{janin2003icsi}, and Mixer6.
AMI and ICSI provide multi-party meeting speech with both ASR and
diarization supervision, while Mixer6 adds distant-microphone interviews
from CHiME-8~\cite{cornell2024chime8} and provides ASR supervision only.
ICSI training meetings with more than $K=8$ speakers are excluded because
they exceed the available diarization slots.
For auxiliary tasks - diarization pretraining, we use $\approx$200 hours of supervised
data, combining AMI with simulated 3--5 speaker conversations generated
from LibriSpeech~\cite{panayotov2015librispeech} using the NeMo
simulator~\cite{park2023property}. This supervision trains the diarization
module to extract speaker activity and speaker-aware representations from
the frozen AudioLLM features.

\begin{table}[t]
\centering
\vspace{-0.2cm}
\caption{Detailed statistics of training chunk. }
\label{tab:train_chunks}
\scriptsize
\setlength{\tabcolsep}{3pt}
\renewcommand{\arraystretch}{0.9}
\begin{tabular}{lrrrrr}
\toprule
\textbf{Corpus} & \textbf{Chunks} & \textbf{Sess.} &
\textbf{Turn/Ch.} & \textbf{Spk/Ch.} & \textbf{Hours} \\
\midrule
AMI    & 14,532 & 136 & 4.42 & 4.00 & 58.22 \\
ICSI   &  8,847 &  49 & 8.82 & 6.21 & 42.80 \\
Mixer6 &  9,533 & 189 & 3.14 & 1.00 & 41.84 \\
\midrule
\textbf{Total} & \textbf{32,912} & \textbf{374} &
\textbf{5.23} & \textbf{3.73} & \textbf{142.86} \\
\bottomrule
\end{tabular}
\vspace{-0.15cm}
\end{table}

\noindent\textbf{Evaluation sets:} We evaluate on held-out AMI, ICSI, and on unseen
LibriCSS~\cite{chen2020continuous} and NOTSOFAR-1~\cite{vinnikov2024notsofar}.
For AMI, we follow the standard pyannote full-corpus partition
~\cite{pyannote_ami_setup} and report both IHM-Mix and SDM conditions.

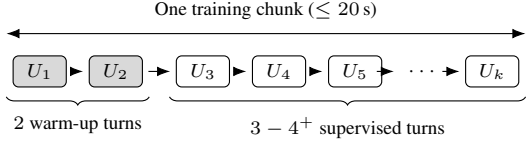
\begin{figure}[t]
\centering
\resizebox{0.8\columnwidth}{!}{%
\begin{tikzpicture}[
    font=\scriptsize,
    turn/.style={
        draw,
        rounded corners=2pt,
        minimum width=0.68cm,
        minimum height=0.40cm,
        inner sep=1pt,
        align=center
    },
    warm/.style={
        turn,
        fill=gray!30
    },
    sup/.style={
        turn,
        fill=white
    },
    arrow/.style={
        -{Latex},
        thin,
        shorten >=2pt,
        shorten <=2pt
    }
]

\node[warm] (u1) {$U_1$};
\node[warm, right=0.28cm of u1] (u2) {$U_2$};

\node[sup, right=0.42cm of u2] (u3) {$U_3$};
\node[sup, right=0.28cm of u3] (u4) {$U_4$};
\node[sup, right=0.28cm of u4] (u5) {$U_5$};

\node[right=0.22cm of u5] (dots) {$\cdots$};
\node[sup, right=0.22cm of dots] (uk) {$U_k$};

\draw[arrow] (u1.east) -- (u2.west);
\draw[arrow] (u2.east) -- (u3.west);
\draw[arrow] (u3.east) -- (u4.west);
\draw[arrow] (u4.east) -- (u5.west);
\draw[arrow] (u5.east) -- (dots.west);
\draw[arrow] (dots.east) -- (uk.west);

\draw[{Latex}-{Latex}, thin]
    ($(u1.north west)+(-0.08cm,0.28cm)$) --
    ($(uk.north east)+(0.08cm,0.28cm)$)
    node[midway, above=2pt]
    {One training chunk ($\leq 20$\,s)};

\draw[
    thin,
    decorate,
    decoration={brace,mirror,amplitude=3pt}
]
    ($(u1.south west)+(-0.08cm,-0.12cm)$) --
    ($(u2.south east)+(0.08cm,-0.12cm)$)
    node[midway,below=4pt] {$2$ warm-up turns};

\draw[
    thin,
    decorate,
    decoration={brace,mirror,amplitude=3pt}
]
    ($(u3.south west)+(-0.08cm,-0.12cm)$) --
    ($(uk.south east)+(0.08cm,-0.12cm)$)
    node[midway,below=4pt] {$3-4^+$ supervised turns};

\end{tikzpicture}%
}
\vspace{-0.3cm}
\caption{Structure of a training chunk. Each chunk spans at most 20\,s and preserves complete speaker turns. The first two turns warm up the memory states, while the remaining 4+ turns contribute to the ASR loss.}
\label{fig:chunking}
\vspace{-0.2cm}
\end{figure}

\subsection{Training Data Preparation}
\label{ssec:data_prep}

We use Qwen2.5-Omni-7B~\cite{xu2025qwen25} as the AudioLLM backbone and keep its audio encoder frozen. We extract $H^{\mathrm{mid}}$ from layer $\ell=15$ of the 32-layer encoder, selected by a layer-wise speaker probe (94.3\% cross-meeting speaker identification accuracy; 10.7\% EER), and cache both $H^{\mathrm{mid}}$ and the final semantic representation $H^{\mathrm{sem}}$.

For efficient learning, long recordings are divided at speaker-turn boundaries into training chunks of at most 20s. As illustrated in Figure~\ref{fig:chunking}, the first two turns warm up the speaker and conversational memories, while the remaining 3-4+ turns contribute to the ASR objective. Overlapping speech is retained through the multi-label speaker activity representation.

\subsection{Training Strategy}
\label{sec:training_strat}

We initialize \stamp{} with the pretrained diarization module and first
perform one epoch of diarization-only warm-up on AMI and ICSI. The remaining
modules are frozen during this stage.

We then jointly optimize the diarization head, speaker-temporal FiLM,
memory Q-Former, and LoRA adapters on the Qwen thinker. For AMI and ICSI, the overall loss $\mathcal{L}
= \mathcal{L}_{\mathrm{LM}} + \mathcal{L}_{\mathrm{diar}}$ 
while Mixer6 contributes only $\mathcal{L}_{\mathrm{LM}}$. The LM objective
is teacher-forced causal cross entropy over transcript tokens.

To bridge reference-guided training and predicted-diarization inference,
we gradually reduce the probability of using reference speaker activity
from 1 to 0, after a 100-step warm-up, over the following 1,200 optimization
steps. 
We jointly optimize the diarization head, speaker-temporal FiLM, memory Q-Former, and LoRA adapters applied to the Qwen thinker, while keeping the AudioLLM encoder and base LLM weights frozen.
We train FiLM and the Q-Former with a learning rate of $10^{-4}$, LoRA ($r=8$, $\alpha=16$) with
$2\times10^{-5}$, and the diarization head with $2\times10^{-6}$.

\noindent\textbf{Diarization pretraining.}
We pretrain the diarization module on AMI and 130 hours of simulated LibriSpeech conversations. 
We use the hybrid Sortformer objective~\cite{park2025sortformer}, $\mathcal{L}_{\mathrm{diar}} = \alpha\mathcal{L}_{\mathrm{sort}} +(1-\alpha)\mathcal{L}_{\mathrm{PIL}}$, with $\alpha=0.25$.

\subsection{Baselines}

We compare \stamp{} with Qwen2.5-Omni-7B as the underlying AudioLLM, Whisper large-v3 with pyannote as a cascaded ASR--diarization baseline, and TagSpeech-AMI as a closely related speaker-temporal AudioLLM.
TagSpeech uses separately trained Zipformer-based models to provide speaker-temporal information, whereas \stamp{} derives speaker activity and representations directly from the AudioLLM's internal features,
without an external speaker encoder or diarization frontend. Unlike TagSpeech, \stamp{} further maintains both speaker and conversational memories across turns.

\subsection{Evaluation Measures}
\label{sec:eval}

We evaluate transcription using WER, concatenated minimum-permutation WER (cpWER), and time-constrained cpWER (tcpWER@5), with cpWER and tcpWER@5 as the primary speaker-attributed ASR metrics. Diarization is evaluated using diarization error rate (DER) with a 0.25s collar. Before ASR scoring, reference and hypothesis transcripts are lowercased and normalized for punctuation and whitespace using the same procedure.
As for the session-level evaluation we additionally report MeetEval's greedy diarization-invariant cpWER (gDI-cpWER)~\cite{von2025word}, which allows hypothesis segmentation to differ from the reference and uses greedy assignment for diarization-invariant scoring over full recording.

\section{Results and Analysis}

\begin{table*}[t]
\centering
\caption{Reported \textbf{cpWER / tcpWER@5 (WER)} ($\downarrow$) under reference-defined and VAD-based segmentation. 
Bold marks the best comparable
cpWER/tcpWER@5; for \stamp{}, \textit{ref.} and \textit{pred.} means reference and predicted speaker activity, respectively, and italicized results use reference speaker activity. }
\label{tab:main_asr}
\vspace{0.05cm}
\scriptsize
\setlength{\tabcolsep}{5.5pt}
\renewcommand{\arraystretch}{1.0}

\begin{tabular}{lccccc}
\toprule
\textbf{Model}
& \textbf{AMI-IHM}
& \textbf{AMI-SDM}
& \textbf{ICSI}
& \textbf{LibriCSS}
& \textbf{NOTSOFAR-1} \\
\midrule


\multicolumn{6}{c}{\textit{Reference-defined segments ($\approx$20\,s)}} \\
\midrule




Whisper-LV3 + pyannote
& 57.2/58.6 (55.1)
& 61.6/63.1 (58.7)
& 58.2/60.7 (56.0)
& 25.5/26.0 (21.2)
& \textbf{62.9/64.0} (52.1) \\
Qwen2.5-Omni-7B
& 58.5/60.0 (57.7)
& 66.5/68.1 (65.6)
& \textbf{46.9/47.5} (45.6)
& \textbf{21.6/21.8}(21.4)
& 67.0/67.9 (65.2) \\

TagSpeech-AMI$^{\star}$
& 54.3/57.2 (38.2)
& \textbf{55.3/57.8} (40.6)
& 53.4/55.5 (31.2)
& 63.2/64.8(32.7)
& 81.0/83.3(62.3) \\

\rowcolor{stampcolor}
\textit{\stamp{} (ref.)}
& \textit{25.3/28.3 (25.0)}
& \textit{38.8/41.5 (38.6)}
& \textit{20.4/21.2 (20.4)}
& \textit{18.5/18.8 (18.5)}
& \textit{52.6/53.7 (52.2)} \\

\rowcolor{stampcolor}
\stamp{} (pred.)
& \textbf{51.5/53.0} (44.1)
& 64.6/66.4 (57.0)
& 47.7/49.3 (42.5)
& 53.3/55.9 (35.6)
& 76.6/79.8/(67.6) \\

\midrule
\multicolumn{6}{c}{\textit{VAD-based segments ($\approx$20\,s)}} \\
\midrule

Qwen2.5-Omni-7B
& 82.8/84.7 (55.8)
& 89.0/91.0 (66.2)
& 82.8/83.9 (44.6)
& 94.4/95.1 (33.8)
& 101.5/102.0 (61.3) \\

TagSpeech-AMI$^{\star}$
& 69.8/72.7 (53.8)
& \textbf{71.7/74.4} (56.6)
& 66.3/69.0 (44.7)
& 80.4/83.0 (47.8)
& 83.6/86.3 (65.2) \\

\rowcolor{stampcolor}
\stamp{} (pred.)
& \textbf{61.8/64.6} (53.1)
& 73.3/76.6 (65.2)
& \textbf{58.2/60.4} (50.8)
& \textbf{69.0/71.9}(49.8)
& \textbf{78.1/81.3} (68.7) \\

\bottomrule
\end{tabular}

\vspace{0.06cm}
\parbox{\textwidth}{\scriptsize $^{\star}$\,TagSpeech-AMI is evaluated on our own
segments and references, not an exact replication of \cite{huo2026tagspeech}: we do not
remove silence or edit the audio, and apply none of the preprocessing used there.}
\vspace{-0.2cm}
\end{table*}

\paragraph*{Multi-Speaker ASR}
Table~\ref{tab:main_asr} reports multi-speaker ASR under reference-defined and VAD-based segmentation. With reference-defined segments, \stamp{} with predicted speaker activity achieves the best comparable cpWER on AMI-IHM and remains competitive on ICSI. Using reference speaker activity substantially reduces error across all test sets, indicating that speaker activity estimation remains an important bottleneck.

Under the more challenging VAD-based setting, \stamp{} achieves the best cpWER on four of five test sets: AMI-IHM, ICSI, LibriCSS, and NOTSOFAR-1. TagSpeech performs slightly better on AMI-SDM (71.7\% vs.\ 73.3\%). Overall, the degradation from reference to predicted speaker activity and from reference-defined to VAD segmentation highlights the importance of robust speaker-temporal estimation for multi-speaker ASR.

\textit{Session-level ASR} Table~\ref{tab:session-asr} reports full-recording gDI-cpWER. \stamp{} outperforms TagSpeech on AMI-IHM, AMI-SDM, and LibriCSS, while the two systems perform similarly on ICSI and NOTSOFAR-1.
Interestingly, on AMI-SDM, TagSpeech performs slightly better in chunk-level cpWER (71.7 vs.\ 73.3), while \stamp{} achieves lower full-session gDI-cpWER (49.7 vs.\ 54.8). This suggests that, when differences in diarization segmentation and assignment are reduced by diarization-invariant scoring, \stamp{} provides stronger full-session transcription, although further exploration is needed to understand the gain.

\begin{table*}[t]
\centering
\caption{Inference-time \stamp{} component analysis using the same jointly trained checkpoint. 
ST = speaker-temporal conditioning, SM = speaker memory, and CM = conversation memory; 
All denotes ST+SM+CM. Results are \textbf{cpWER / tcpWER@5 (WER)} on reference-defined 
segments. Best results are bold; dev sets show consistent trends.}

\label{tab:stamp-components}
\vspace{0.05cm}
\scriptsize
\setlength{\tabcolsep}{4.2pt}
\renewcommand{\arraystretch}{1.02}

\begin{tabular}{lccccc}
\toprule
\textbf{Inference configuration} &
\textbf{AMI-IHM} &
\textbf{AMI-SDM} &
\textbf{ICSI} &
\textbf{LibriCSS} &
\textbf{NOTSOFAR-1} \\
\midrule

\multicolumn{6}{c}{\textit{Reference speaker activity}} \\
\midrule

\rowcolor{stampcolor}
All (ST+SM+CM)
& 25.3 / 28.3 (25.0)
& 38.8 / 41.5 (38.6)
& \textbf{20.4} / \textbf{21.2} (20.4)
& 18.5 / 18.8 (18.5)
& 52.6 / 53.7 (52.2) \\

ST
& 25.8 / 29.1 (25.5)
& \textbf{33.7} / \textbf{37.2} (33.5)
& 23.7 / 24.5 (23.7)
& \textbf{13.9} / \textbf{14.1} (13.9)
& \textbf{47.0} / \textbf{48.0} (46.7) \\

ST+SM
& 25.2 / 28.3 (24.9)
& 37.0 / 39.9 (36.8)
& 20.9 / 21.6 (21.0)
& 17.1 / 17.4 (17.1)
& 50.7 / 51.8 (50.2) \\

ST+CM
& \textbf{25.1} / \textbf{28.0} (24.8)
& 38.4 / 41.0 (38.2)
& 20.4 / 21.2 (20.3)
& 19.7 / 20.0 (19.6)
& 52.1 / 53.2 (51.6) \\

\midrule
\multicolumn{6}{c}{\textit{Predicted speaker activity}} \\
\midrule

\rowcolor{stampcolor}
All (ST+SM+CM)
& \textbf{51.5} / \textbf{53.0} (44.1)
& 64.6 / 66.4 (57.0)
& \textbf{47.7} / \textbf{49.3} (42.5)
& 53.3 / 55.9 (35.6)
& 76.6 / 79.8 (67.6) \\

ST
& 53.3 / 54.5 (44.9)
& \textbf{61.6} / \textbf{63.0} (53.4)
& 51.8 / 52.8 (46.7)
& \textbf{51.1} / \textbf{53.3} (32.0)
& \textbf{75.9} / \textbf{78.1} (64.1) \\

ST+SM
& 52.0 / 53.4 (44.3)
& 62.0 / 63.9 (54.5)
& 48.1 / 50.1 (43.5)
& 52.1 / 55.1 (35.3)
& 75.4/ 79.1 (66.2) \\

ST+CM
& \textit{51.5 }/ \textit{53.1} (43.5)
& 63.6 / 65.5 (56.4)
& 48.0 / 49.3 (43.4)
& 54.5 / 57.2 (38.7)
& 76.2 / 79.8 (67.4) \\

\bottomrule
\end{tabular}
\vspace{-0.2cm}
\end{table*}

\paragraph*{When Does Each \stamp{} Component Help?}
Table~\ref{tab:stamp-components} analyzes different inference configurations of the same jointly trained \stamp{}-ASR checkpoint. Focusing on predicted speaker activity, two patterns emerge. On the natural close-talk meeting sets, AMI-IHM and ICSI, using all \stamp{} components reduces cpWER from 53.3 to 51.5 and from 51.8 to 47.7, respectively. In contrast, speaker-temporal conditioning alone performs best on AMI-SDM and LibriCSS, where adding memory consistently increases error. NOTSOFAR-1 shows only a marginal cpWER gain from speaker memory, while speaker-temporal conditioning gives better tcpWER@5 and WER.

These results suggest that \stamp{}'s memory is more useful when persistent speaker and conversational context is reliable, whereas current speaker-temporal evidence can be sufficient for more challenging
far-field or cross-domain conditions, a clear evident can be seen when comparing the two versions of the AMI. Importantly, these configurations require no retraining from the architectural point-of-view and can be selectively use these capabilities learned by the same \stamp{} model.





\begin{table}[t]
\centering
\caption{Full-session gDI-cpWER (\%; $\downarrow$) using VAD segmentation and
scored over complete recordings.}
\label{tab:session-asr}
\scriptsize
\setlength{\tabcolsep}{3.5pt}
\renewcommand{\arraystretch}{0.9}
\begin{tabular}{lccccc}
\toprule
\textbf{Model} &
\textbf{AMI-IHM} &
\textbf{AMI-SDM} &
\textbf{ICSI} &
\textbf{LibriCSS} &
\textbf{NSF-1} \\
\midrule
TagSpeech-AMI
& 52.1 & 54.8 & \textbf{44.8} & 48.9 & \textbf{65.1} \\
\stamp{}
& \textbf{37.6} & \textbf{49.7} & 45.0 & \textbf{41.5} & 65.5 \\
\bottomrule
\end{tabular}
\vspace{-0.3cm}
\end{table}

\begin{table}[t]
\centering
\caption{Diarization error rate (DER) and speaker confusion (Conf.), with a
0.25\,s collar. $\mathcal{D}_{\phi}$ reports mean $\pm$ s.d.\ over three
seeds. $^{\ddagger}$Sortformer covers 6/16 AMI meetings; parentheses show
$\mathcal{D}_{\phi}$ on the same subset. $^{*}$TagSpeech uses 20--25\,s
chunk-local scoring, which does not penalize cross-chunk speaker identity.}
\label{tab:diar-bench}
\vspace{0.05cm}
\scriptsize
\setlength{\tabcolsep}{2.5pt}
\renewcommand{\arraystretch}{0.8}
\begin{tabular}{lcccccccc}
\toprule
 & \multicolumn{2}{c}{$\mathcal{D}_{\phi}$} &
   \multicolumn{2}{c}{Sortformer} &
   \multicolumn{2}{c}{pyannote} &
   \multicolumn{2}{c}{TagSpeech$^{*}$} \\
\cmidrule(lr){2-3}\cmidrule(lr){4-5}\cmidrule(lr){6-7}\cmidrule(lr){8-9}
\textbf{Test} & DER & Conf. & DER & Conf. & DER & Conf. & DER & Conf. \\
\midrule
AMI-IHM    & $36.6{\pm}3.8$ (32.3) & 17.4 & $29.0^{\ddagger}$ & 7.7  & \textbf{12.3} & 3.7  & 39.9 & 7.8 \\
AMI-SDM    & $51.4{\pm}4.8$ (49.3) & 17.6 & $35.3^{\ddagger}$ & 13.7 & \textbf{15.4} & 5.2  & 37.8 & 8.3 \\
ICSI       & $57.0{\pm}1.9$        & 28.9 & --   & --   & \textbf{25.1} & 3.1  & 34.7 & 11.2 \\
LibriCSS   & $56.7{\pm}1.1$        & 45.2 & 46.7 & 34.5 & \textbf{11.3} & 3.7  & 33.8 & 15.6 \\
NOTSOFAR-1 & $40.5{\pm}0.8$        & 27.8 & 22.5 & 10.2 & \textbf{19.9} & 11.1 & 54.2 & 11.7 \\
\bottomrule
\end{tabular}
\vspace{-0.2cm}
\end{table}



\paragraph*{Auxiliary Diarization Analysis.}
Table~\ref{tab:diar-bench} evaluates the lightweight diarization module that provides speaker-temporal cues to \stamp{}. Our primary $\mathcal{D}_{\phi}$ results use full-recording speaker identities and therefore penalize speaker confusion across the entire session. In contrast, TagSpeech operates on 20--25\,s chunks with chunk-local speaker identities, which does not penalize cross-chunk identity errors. Under the same chunk-local protocol, $\mathcal{D}_{\phi}$ achieves 28.4\% DER on AMI-IHM, compared with 39.9\% for TagSpeech, showing that the \stamp{} diarization component provides effective local speaker-temporal cues.

Performance drops when $\mathcal{D}_{\phi}$ is evaluated over complete recordings (36.6\% DER on AMI-IHM), with speaker confusion becoming the dominant error on many-speaker conditions such as ICSI and LibriCSS. 
This suggests that maintaining speaker identity over long recordings is one of the main limitation than local speaker activity estimation.


\paragraph*{Key findings}
Our results show \textit{three main} trends. First, \stamp{} performs strongly across diverse meeting conditions, particularly under automatic segmentation.
Second, speaker-temporal conditioning and memory provide complementary benefits depending on the conversational and acoustic condition. 
Finally, the large gap between reference and predicted speaker activity shows that the robust speaker tracking remains a key bottleneck.


\section{Conclusion}

We introduced \textbf{\stamp{}}, a lightweight framework that augments a pretrained AudioLLM with speaker-temporal anchoring and fixed-size speaker and conversational memories for multi-speaker ASR. Our experiments show that speaker-temporal conditioning and memory provide complementary benefits across different conversational conditions, while predicted speaker activity remains a key source of error. These findings highlight the importance of jointly modeling current speaker evidence and persistent conversational context for
robust multi-speaker ASR.



\bibliographystyle{IEEEbib}
\bibliography{strings,refs}

@article{taherian2022multi,
  author  = {Hassan Taherian and Ke Tan and DeLiang Wang},
  title   = {Multi-Channel Talker-Independent Speaker Separation through Location-Based Training},
  journal = {IEEE/ACM TASLP},
  year    = {2022}
}

@inproceedings{chen2020continuous,
  author    = {Zhuo Chen and Takuya Yoshioka and Liang Lu and Tianyan Zhou and Zhong Meng and Yi Luo and Jian Wu and Xiong Xiao and Jinyu Li},
  title     = {Continuous Speech Separation: Dataset and Analysis},
  booktitle = {Proc. ICASSP},
  year      = {2020}
}

@inproceedings{kanda2020serialized,
  author    = {Naoyuki Kanda and Yashesh Gaur and Xiaofei Wang and Zhong Meng and Takuya Yoshioka},
  title     = {Serialized Output Training for End-to-End Overlapped Speech Recognition},
  booktitle = {Proc. Interspeech},
  year      = {2020}
}

@inproceedings{kanda2022streaming,
  author    = {Naoyuki Kanda and Jian Wu and Yu Wu and Xiong Xiao and Zhong Meng and Xiaofei Wang and Yashesh Gaur and Zhuo Chen and Jinyu Li and Takuya Yoshioka},
  title     = {Streaming Multi-Talker ASR with Token-Level Serialized Output Training},
  booktitle = {Proc. Interspeech},
  year      = {2022}
}

@article{raj2023surt,
  author  = {Desh Raj and Daniel Povey and Sanjeev Khudanpur},
  title   = {{SURT} 2.0: Advances in Transducer-Based Multi-Talker Speech Recognition},
  journal = {IEEE/ACM TASLP},
  year    = {2023}
}

@inproceedings{park2025sortformer,
  author    = {Taejin Park and others},
  title     = {Sortformer: A Novel Approach for Permutation-Resolved Speaker Supervision in Speech-to-Text Systems},
  booktitle = {Proc. ICML},
  year      = {2025}
}

@inproceedings{kanda20_interspeech,
  author    = {Naoyuki Kanda and Yashesh Gaur and Xiaofei Wang and Zhong Meng and Zhuo Chen and Tianyan Zhou and Takuya Yoshioka},
  title     = {Joint Speaker Counting, Speech Recognition, and Speaker Identification for Overlapped Speech of Any Number of Speakers},
  booktitle = {Proc. Interspeech},
  year      = {2020}
}

@inproceedings{kanda2022transcribe,
  author    = {Naoyuki Kanda and Xiong Xiao and Yashesh Gaur and Xiaofei Wang and Zhong Meng and Zhuo Chen and Takuya Yoshioka},
  title     = {Transcribe-to-Diarize: Neural Speaker Diarization for Unlimited Number of Speakers Using End-to-End Speaker-Attributed {ASR}},
  booktitle = {Proc. ICASSP},
  year      = {2022}
}

@inproceedings{delcroix2018single,
  author    = {Marc Delcroix and Katerina Zmolikova and Keisuke Kinoshita and Atsunori Ogawa and Tomohiro Nakatani},
  title     = {Single Channel Target Speaker Extraction and Recognition with Speaker Beam},
  booktitle = {Proc. ICASSP},
  year      = {2018}
}

@article{polok2026dicow,
  author  = {Alexander Polok and others},
  title   = {{DiCoW}: Diarization-Conditioned Whisper for Target Speaker Automatic Speech Recognition},
  journal = {Comput. Speech Lang.},
  year    = {2026}
}

@inproceedings{shi2024advancing,
  author    = {Mohan Shi and Zengrui Jin and Yaoxun Xu and Yong Xu and Shi-Xiong Zhang and Kun Wei and Yiwen Shao and Chunlei Zhang and Dong Yu},
  title     = {Advancing Multi-Talker ASR Performance with Large Language Models},
  booktitle = {Proc. SLT},
  year      = {2024}
}

@inproceedings{meng2025large,
  author    = {Lingwei Meng and Shujie Hu and Jiawen Kang and Zhaoqing Li and Yuejiao Wang and Wenxuan Wu and Xixin Wu and Xunying Liu and Helen Meng},
  title     = {Large Language Model Can Transcribe Speech in Multi-Talker Scenarios with Versatile Instructions},
  booktitle = {Proc. ICASSP},
  year      = {2025}
}

@inproceedings{yin2026speakerlm,
  author    = {Han Yin and Yafeng Chen and Chong Deng and Luyao Cheng and Hui Wang and Chao-Hong Tan and Qian Chen and Wen Wang and Xiangang Li},
  title     = {{SpeakerLM}: End-to-End Versatile Speaker Diarization and Recognition with Multimodal Large Language Models},
  booktitle = {Proc. AAAI},
  year      = {2026}
}

@inproceedings{huo2026tagspeech,
  author    = {Mingyue Huo and Yiwen Shao and Yuheng Zhang},
  title     = {{TagSpeech}: End-to-End Multi-Speaker ASR and Diarization with Fine-Grained Temporal Grounding},
  booktitle = {Proc. ACL},
  year      = {2026}
}

@misc{lin2026speakerreasoner,
  author = {Zhennan Lin and Shuai Wang and Zhaokai Sun and Pengyuan Xie and Chuan Xie and Jie Liu and Qiang Zhang and Lei Xie},
  title  = {Speaker-Reasoner: Scaling Interaction Turns and Reasoning Patterns for Timestamped Speaker-Attributed ASR},
  note   = {arXiv},
  year   = {2026}
}

@inproceedings{shi2026train,
  author    = {Mohan Shi and Xiong Xiao and Ruchao Fan and Shaoshi Ling and Jinyu Li},
  title     = {Train Short, Infer Long: Speech-LLM Enables Zero-Shot Streamable Joint ASR and Diarization on Long Audio},
  booktitle = {Proc. ICASSP},
  year      = {2026}
}

@article{peng2026g,
  author  = {Jing Peng and Ziyi Chen and Haoyu Li and Yucheng Wang and Duo Ma and Mengtian Li and Yunfan Du and Dezhu Xu and Kai Yu and Shuai Wang},
  title   = {{G-STAR}: End-to-End Global Speaker-Tracking Attributed Recognition},
  journal = {arXiv},
  year    = {2026}
}

@inproceedings{hori20_interspeech,
  author    = {Takaaki Hori and Niko Moritz and Chiori Hori and Jonathan Le Roux},
  title     = {Transformer-Based Long-Context End-to-End Speech Recognition},
  booktitle = {Proc. Interspeech},
  year      = {2020}
}

@inproceedings{dai2019transformer,
  author    = {Zihang Dai and Zhilin Yang and Yiming Yang and Jaime G. Carbonell and Quoc V. Le and Ruslan Salakhutdinov},
  title     = {Transformer-XL: Attentive Language Models Beyond a Fixed-Length Context},
  booktitle = {Proc. ACL},
  year      = {2019}
}

@inproceedings{bulatov2022recurrent,
  author    = {Aydar Bulatov and Yuri Kuratov and Mikhail Burtsev},
  title     = {Recurrent Memory Transformer},
  booktitle = {Proc. NeurIPS},
  year      = {2022}
}

@inproceedings{perez2018film,
  title={Film: Visual reasoning with a general conditioning layer},
  author={Perez, Ethan and Strub, Florian and De Vries, Harm and Dumoulin, Vincent and Courville, Aaron},
  booktitle={Proceedings of the AAAI conference on artificial intelligence},
  volume={32},
  number={1},
  year={2018}
}

@misc{xu2025qwen25,
      title={Qwen2.5-Omni Technical Report}, 
      author={Jin Xu and others},
      year={2025},
      eprint={2503.20215},
      archivePrefix={arXiv},
      primaryClass={cs.CL},
      url={https://arxiv.org/abs/2503.20215}, 
}

@inproceedings{carletta2005ami,
  author    = {Jean Carletta and others},
  title     = {The {AMI} Meeting Corpus: A Pre-Announcement},
  booktitle = {Proc. MLMI},
  year      = {2005}
}

@inproceedings{janin2003icsi,
  author    = {Adam Janin and others},
  title     = {The {ICSI} Meeting Corpus},
  booktitle = {Proc. ICASSP},
  year      = {2003}
}

@article{cornell2024chime8,
  author  = {Samuele Cornell and others},
  title   = {The {CHiME}-8 {DASR} Challenge for Generalizable and Array-Agnostic Distant Automatic Speech Recognition and Diarization},
  journal = {arXiv preprint arXiv:2407.16447},
  year    = {2024}
}

@inproceedings{vinnikov2024notsofar,
  author    = {Alon Vinnikov and others},
  title     = {{NOTSOFAR}-1 Challenge: New Datasets, Baseline, and Tasks for Distant Meeting Transcription},
  booktitle = {Proc. CHiME},
  year      = {2024}
}

@inproceedings{panayotov2015librispeech,
  author    = {Vassil Panayotov and Guoguo Chen and Daniel Povey and Sanjeev Khudanpur},
  title     = {{LibriSpeech}: An {ASR} Corpus Based on Public Domain Audio Books},
  booktitle = {Proc. ICASSP},
  year      = {2015}
}

@inproceedings{park2023property,
  author    = {Tae Jin Park and others},
  title     = {Property-Aware Multi-Speaker Data Simulation: A Probabilistic Modelling Technique for Synthetic Data Generation},
  booktitle = {Proc. CHiME},
  year      = {2023}
}

@misc{pyannote_ami_setup,
  title        = {{AMI} Diarization Setup},
  author       = {{pyannote}},
  howpublished = {\url{https://github.com/pyannote/AMI-diarization-setup}},
  year         = {2021}
}

@article{von2025word,
  title={Word error rate definitions and algorithms for long-form multi-talker speech recognition},
  author={von Neumann, Thilo and Boeddeker, Christoph and Delcroix, Marc and Haeb-Umbach, Reinhold},
  journal={IEEE Transactions on Audio, Speech and Language Processing},
  year={2025},
  publisher={IEEE}
}

\end{document}